\documentclass[pra,a4paper,superscriptaddress,twocolumn,amsmath,amssymb,floatfix,amstex]{revtex4-1}%
\usepackage{graphicx,graphics,rotating}	% Include figure files
\usepackage{dcolumn}
\usepackage{bm}
\usepackage{enumerate}
\usepackage{natbib}
\usepackage{hyperref}
\usepackage{amsmath}
\usepackage{mathtools}
\usepackage{varwidth}
\usepackage{hyperref}
\usepackage{mathptmx}
\usepackage{color}
\usepackage{soul}
\usepackage{mathtools}
\usepackage{leftidx}
\begin{document}

\title{Quantum repeater without Bell measurements in double quantum dot systems}

\author{Xiao-Feng Yi}
\email[Corresponding author: ]{xiaofengyi@whu.edu.cn}
%\affiliation{ School of physics and Technology,Wuhan University,Wuhan 430072,China}
\author{Peng Xu}
%\affiliation{School of physics and Technology,Wuhan University,Wuhan 430072,China}
\author{Qi Yao}
\affiliation{School of physics and Technology,Wuhan University,Wuhan 430072,China}
\author{Xianfu Quan}
\affiliation{State key Laboratory of Magnetic Resonance and Atomic and Molecular physics,Wuhan Institute of Physics and Mathematics,Chinese Academy of Sciences,Wuhan 430071,China}
\affiliation{University of Chinese Academy of Science,Bejing 10079,China}

\date{\today}%
%%%
\begin{abstract}
We propose a Bell measurement free scheme to implement a quantum repeater in GaAs/AlGaAs double quantum dot systems. We prove that four pairs of double quantum dots compose an entanglement unit, given that the initial state is singlet states. Our scheme differs from the famous Duan-Lukin-Cirac-Zoller (DLCZ) protocol in that the Bell measurements are unnecessary for the entanglement swapping, which provides great advantages and conveniences in experimental implementations. Our scheme significantly improves the success probability of quantum repeaters based on solid state quantum devices.
\end{abstract}

\maketitle

%%%%%%%%%%%%%%%%%%%%%%%%%%%%%%%%%%%%%%%%%%%%%%%%%%%%%%%%%%%%%%%%%%%%%%%%%%%%%%%%%%%%%%%%%%%%%%%%%%%%%%%%
\section{\label{sec:levell}INTRODUCTION}

Quantum repeater is a basic building block in quantum communication, quantum computing, and quantum teleportation. After the original ideal of the quantum repeater by Briegel {\it el al.}~\cite{PhysRevLett.81.5932} in 1998, Duan {\it el al.}~\cite{Duan2001Long} presented the widely adopted DLCZ scheme, which is based on atomic ensembles and linear optics with many Bell measurements. Shortly, the robustness of a quantum repeater was checked by Zhao {\it el al.}~\cite{PhysRevLett.98.240502} and Jiang {\it el al.}~\cite{PhysRevA.76.012301}. Numerous quantum repeater schemes to further enhance the noise-resistance are followed. Among all these schemes the photons as an information carrier and the Bell measurements are required. Similar to the DLCZ scheme, the probability of the successful entanglement swapping is 50\%~\cite{PhysRevLett.80.3891}. To improve the success probability, entanglement purification is also adopted~\cite{Pan2000Entanglement}.

There are many physical systems serving as candidates to realize quantum information processing~\cite{PhysRevLett.83.4204}, for example, trapped ions, quantum dots (QDs), photons, neutral cold atoms in optical lattice, nitrogen vacancy centers~\cite{PhysRevA.95.052336} in diamond, and superconducting qubits~\cite{Ladd2010Quantum}. Among these systems, we focus on the semiconductor QDs with the aid of optical cavities. The QD is often called artificially atom, which consists of electrons or holes confined in a potential well. Many materials may form QDs~\cite{RevModPhys.79.1217}, such as semiconductor QDs (InAs or GaAs), graphene QDs, and so on. We specifically take the semiconductor QDs into consideration in this paper, which are constructed by heterostructures of GaAs and AlGaAs grown with the molecular beam epitaxy technique~\cite{cho1975molecular}.

QDs have been employed to realize a quantum repeater. In 2014, C. Wang {\it et al.} presented a scheme to construct a quantum repeater based on QDs~\cite{Wang2014Construction}. In 2015, Jianping Wang {\it et al.} achieved scalable entangled photon source with self-assembled QDs in experiment~\cite{PhysRevLett.115.067401}, which gives one a hope to implement a quantum repeater in QDs. However, these schemes are based on photons and Bell measurements. Realizing entanglement between two distant QDs is still absent~\cite{Mcmahon2015Towards, PhysRevB.90.235421}.

\begin{figure}[ht]
\centering
\includegraphics[angle=-90,scale=0.30]{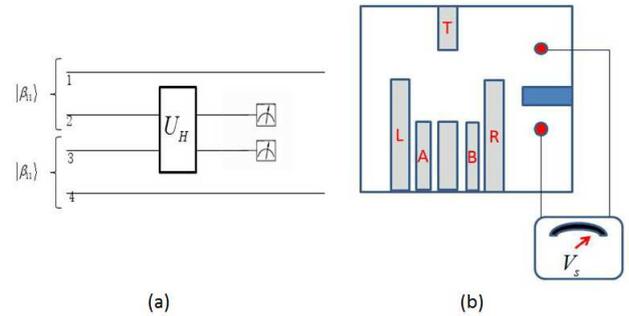}
\caption{(a) Quantum circuit diagram of entanglement swapping. The initial state for the two pairs of DQDs is a singlet state $|\beta_{11}\rangle$. Detecting the QDs $2$ and $3$ after a certain evolution time $t$ establishes the entanglement swapping. (b) A schematic diagram of a double quantum dot. The voltages applied on gates L and R control the number of electrons in the left and right QDs, respectively. The gate T controls the tunnelling strength between the QDs. The depth of potential wells are adjusted by the gates A and B. A quantum point contact (QPC) is used detect the charge states ($n_1, n_2$) with $n_{1, 2}$ denoting the electron in left/right quantum dot.}
\label{fig:figure1}
\end{figure}

In this paper, we present an alternative scheme of a quantum repeater which is based on the double quantum dots (DQDs). Different from previous proposals, our scheme employs the product of local measurements on QD electron spins, instead of the Bell measurements of photons. Such local measurements are easier to implement than Bell measurements in QD experiments. In fact, our scheme is divided into three steps: (i) Preparation of entanglement pairs in DQDs; (ii) Creation of entanglement swapping between 4 QDs, with the help of optical cavities which couple the adjacent DQDs~\cite{RevModPhys.81.865, petta2005coherent, PhysRevLett.85.2392, Chen2006Scheme, Bouwmeester1997Experimental}; (iii) Extending the entanglement distance many times to realize a quantum repeater, as shown in Fig.~\ref{fig:figure1}(a).

%%%%%%%%%%%%%%%%%%%%%%%%%%%%%%%%%%%%%%%%%%%%%%%%%%%%%%%%%%%%%%%%%%%%%%%%%%%%%%%%%%%%%%%%%%%%%%%%%%%%%%%%
\section{\label{sec:level2}Quantum repeater based on DQDs}

\subsection{Preparation of initial $(1,1)$ singlet states}

\begin{figure}
\includegraphics[angle=-90,scale=0.25]{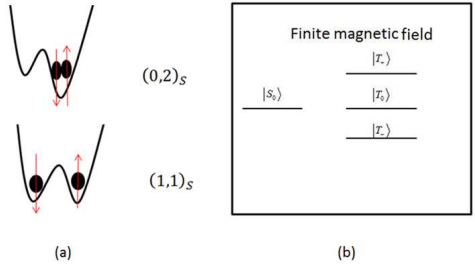}
\caption{(a) Preparation diagram of a singlet state ${(1,1)_S}$ in a DQD. (b) Energy levels of the $(1,1)$ state in a finite magnetic field. The triplet states are split by the Zeeman energy of the field.}
\label{fig:figure2}
\end{figure}

We consider a quantum repeater implemented with many DQDs which are initialized in a $(1,1)$ singlet state (Bell state) for each DQD. Figure~\ref{fig:figure1}(b) is a schematic diagram of a typical DQD. To prepare the initial singlet state in a DQD, as shown in Fig.~\ref{fig:figure3}(a), we consider two electrons in the DQD. The charge configurations includes $(0,2)$ and $(1,1)$, where the two electrons are both in the right QD and each in a QD, respectively. For the $(0,2)$ charge configuration with a large bias (detuning) $\Delta$, the ground state of the DQD is a singlet state $(0,2)_S$ due to the Pauli exclusion principle, as shown in Fig.~\ref{fig:figure2}(a). For the $(1,1)$ charge configuration at zero bias (detuning) $\Delta$, two spin states, a singlet state $(1,1)_S$ and three triplet states $(1,1)_T$, are possible. The bias $\Delta$ can be adjust in experiments by the gate voltages on gate L and gate R. By adiabatically and appropriately lowing the bias $\Delta$ in a magnetic field which splits the triplet states of the electrons as shown in Fig.~\ref{fig:figure2}(b), one reaches the singlet state where the two electron spins in each QD are entangled. This process has been successfully demonstrated in experiments~\cite{petta2005coherent} and the probability of $(1,1)_S$ can be detected by employing a QPC sensor.

\subsection{\label{sec:level2}Entanglement swapping in a pair of DQDs}

\begin{figure}[ht]
\centering
\includegraphics[angle=-90,scale=0.30]{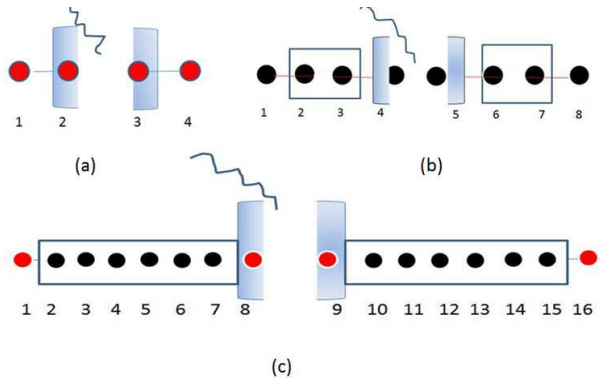}
\caption{(a) Entanglement swapping in two pairs of DQDs. The QDs 2 and 3 are coupled within a cavity. Entanglement is established between QD 1 and QD 8 (b) and between QD 1 and QD 16 (c).}
\label{fig:figure3}
\end{figure}

After the step of preparing many $(1,1)_S$ DQD singlet states, which means we have many locally entangled electron spin pairs, we need to extend the entanglement distance. As a second step, we extend the entanglement distance from DQD to a pair of DQD, i.e., building the entanglement between the electron spins in QD 1 and QD 4. We choose the well-known entanglement swapping method which can be achieved by coupling the middle QDs 2 and 3 equally with a single-mode cavity system~\cite{PhysRevLett.85.2392}, as shown in Fig.~\ref{fig:figure3}(a). This is the famous Tavis-Cummings model (TCM)~\cite{PhysRev.170.379}, whose Hamiltonian is:
\begin{equation}
\label{equ1}
H={\omega}a^{\dag}a+\frac{1}{2}\omega_2\sigma_{2Z}+\frac{1}{2}\omega_3\sigma_{3Z} +\sum\limits_{i=2,3}g(a^{\dag}\sigma_i^- + a\sigma_i^+)
\end{equation}
where $\omega$ is the single mode frequency of the cavity, $\omega_{2,3}$ is the Zeeman splitting of the electron spin in QD2 and QD3 (we assume $\omega_2$=$\omega_3$), respectively, and $g$ is the coupling strength between the QDs and the cavity. The annihilation operator of the cavity mode is $a$, and $\sigma _i^ +  = {\left| e \right\rangle _{ii}}$$\left\langle g \right|$ and $\sigma _i^ -  = {\left| {\rm{g}} \right\rangle _{ii}\left\langle e \right|}$, where ${\left| g \right\rangle _i}$ and ${\left| e \right\rangle _i}$ are the ground state and excited state of the $i$th QD, respectively. We have set $\hbar = 1$.

In the interaction picture defined by $H_0 = {\omega}a^{\dag}a+\frac{1}{2}\omega_2\sigma_{2Z} + \frac{1}{2}\omega_3\sigma_{3Z}$, the Hamilton of the system under the rotating wave approximation can be written as:
\begin{equation}
\label{equ2}
{H_i} = g\sum\limits_{i = 2,3} {({e^{ - i\Delta t}}{a^\dag }\sigma _i^ -  + H.C)}
\end{equation}
where $\Delta$ is the detuning between the cavity frequency $\omega$ and the QD transition frequency $\omega _i$. There is no energy exchange between the cavity and the QDs in the case $\Delta \gg g$. By further employing the Nakajiama Transformation~\cite{PhysRevLett.85.2392}, we obtain the effective Hamiltonian
\begin{equation}
\label{equ3}
\begin{split}
H = \lambda [\sum\limits_{i = 2,3} {({{\left| e \right\rangle }_{ii}\left\langle e \right|a{a^\dag }} - {{\left| g \right\rangle }_{ii}\left\langle g \right|{a^\dag }a)}}\\
  + (\sigma _2^ + \sigma _3^ -  + \sigma _3^ + \sigma _2^ - )]
\end{split}
\end{equation}
with $\lambda  = {g^2}/{\Delta }$. Suppose that the cavity is prepared in a vacuum state, the effective Hamiltonian is then reduced to~\cite{PhysRevLett.85.2392}:
\begin{equation}
\label{equ4}
H = \lambda [\sum\limits_{i = 2,3} {{{\left| e \right\rangle }_{ii}\left\langle e \right| + (\sigma _2^ + \sigma _3^ -  + \sigma _2^ - \sigma _3^ + )}} ].
\end{equation}
Under such a Hamiltonian, the QD $i$ and QD $j$ with an initial product state, such as ${\left| {ee} \right\rangle _{ij}}, {\left| {eg} \right\rangle _{ij}}, {\left| {ge} \right\rangle _{ij}}$, and ${\left| {gg} \right\rangle _{ij}}$, becomes after a period time $t$, respectively,
\begin{equation}
\begin{split}
\label{equ6}
\begin{array}{l}
{\left| {eg} \right\rangle _{ij}} \to {e^{ - i\lambda t}}[\cos (\lambda t){\left| {eg} \right\rangle _{ij}} - i\sin (\lambda t){\left| {ge} \right\rangle _{ij}}], \\
{\left| {ee} \right\rangle _{ij}} \to {e^{ - i2\lambda t}}{\left| {ee} \right\rangle _{ij}}, \\
{\left| {gg} \right\rangle _{ij}} \to {\left| {gg} \right\rangle _{ij}}.
\end{array}
\end{split}
\end{equation}

The initial four QDs are in a product state of a pair of singlet states, i.e.,
\begin{equation}
\label{equ5}
\psi_{1,2,3,4} = \psi_{1,2} \otimes \psi_{3,4}
\end{equation}
with
\begin{equation}
\begin{split}
\psi_{1,2} = \frac{1}{{\sqrt 2 }}(\left| {eg} \right\rangle  - \left| {ge} \right\rangle )_{1,2}, \\
\psi_{3,4} = \frac{1}{{\sqrt 2 }}(\left| {eg} \right\rangle  - \left| {ge} \right\rangle )_{3,4}.
\end{split}
\end{equation}
After an evolution time $t$, the whole state of the four QDs evolves into
\begin{equation}
\begin{split}
\label{equ8}
{\left| \psi  \right\rangle _{1,2,3,4}}
&= {\left| {ge} \right\rangle _{1,4}}{e^{ - i\lambda t}}\left[ {\cos \left( {\lambda t} \right){{\left| {eg} \right\rangle }_{2,3}} - i\sin (\lambda t){{\left| {ge} \right\rangle }_{2,3}}} \right]\\
&\quad- {e^{ - i2\lambda t}}{\left| {gg} \right\rangle _{1,4}}{\left| {ee} \right\rangle _{2,3}}- {\left| {ee} \right\rangle _{1,4}}{\left| {gg} \right\rangle _{2,3}}\\
&\quad+ {\left| {eg} \right\rangle _{1,4}}{e^{ - i\lambda t}}\left( {\cos \left( {\lambda t} \right){{\left| {ge} \right\rangle }_{2,3}} - i\sin (\lambda t){{\left| {eg} \right\rangle }_{2,3}}} \right).
\end{split}
\end{equation}
At a special moment $\lambda t = \pi /4$, the above state Eq.~(\ref{equ8}) turns into
\begin{equation}
\label{equ9}
\begin{split}
\left| \psi  \right\rangle _{1,2,3,4}
&= {e^{ - i\frac{\pi }{4}}}\sqrt{2} {\left| {gg} \right\rangle _{1,4}}{\left| {ee} \right\rangle _{2,3}} - {e^{i\frac{\pi }{4}}}\sqrt{2} {\left| {ee} \right\rangle _{1,4}}{\left| {gg} \right\rangle _{2,3}} \\
&\quad+ \frac{{{e^{ - i\pi /4}}}}{{2\sqrt{2}}}
{\left| {eg} \right\rangle _{2,3}}[{\left| {ge} \right\rangle _{1,4}} - i{\left| {eg} \right\rangle _{1,4}}]\\
&\quad + {\left| {ge} \right\rangle _{2,3}}[{\left| {eg} \right\rangle _{1,4}} - i{\left| {ge} \right\rangle _{1,4}}].
\end{split}
\end{equation}
By measuring the state of the QDs 2 and 3, there are four possible outcomes: ${\left| {gg} \right\rangle _{2,3}}$, ${\left| {ee} \right\rangle _{2,3}}$, ${\left| {eg} \right\rangle _{2,3}}$, and ${\left| {ee} \right\rangle _{2,3}}$. For the first two results, ${\left| {gg} \right\rangle _{2,3}}$ and ${\left| {ee} \right\rangle _{2,3}}$, no entanglement swapping occurs, i.e., QDs 1 and 4 are not entangled. However, if we get one of the last two results ${\left| {eg} \right\rangle _{2,3}}$ or ${\left| {ge} \right\rangle _{2,3}}$, the state would collapse into an entangled state between QDs 1 and 4, namely
\begin{equation}
\label{equ10}
\left|\psi\right\rangle _{1,4} = \frac{1}{\sqrt{2}}[{\left| {ge} \right\rangle _{1,4}} - i{\left| {eg}\right\rangle _{1,4}}]
\end{equation}
or
\begin{equation}
\left|\psi\right\rangle _{1,4}^{'} = \frac{1}{\sqrt{2}}[{\left| {eg} \right\rangle _{1,4}} - i{\left| {ge} \right\rangle _{1,4}}]
\end{equation}
where we have ignored the overall phase.

\subsection{\label{sec:level2}Quantum repeater}

For a practical quantum repeater, one needs to extend the distance between the two entangled qubits (QDs here) as long as possible. In our protocol, starting from many singlet states of two adjacent QDs, we have proved that the entangled swapping can be achieved, i.e., QD 1 and QD 4 are entangled pairs and the entanglement distance is doubled. Next, we need to extend further the entanglement distance to 8 QDs, 16 QDs and more~(Fig.\ref{fig:figure3}(b)), until we find a periodicity of the entanglement.

Next, for an 8 QDs unit, there are 3 possible initial entangled states in this system
\begin{equation}\begin{split}
\label{equ11}
\begin{array}{l}
(1).\; {\psi _{\rm{L}}} = {\psi _R} = {\psi _{1,4}}{\rm{ = }}{\psi _{5,8}},\\
(2).\; {\psi _{\rm{L}}} = {\psi _R} = \psi _{1,4}^{'}{\rm{ = }}{\psi _{5,8}},\\
(3).\; {\psi _{\rm{L}}} = {\psi _{1,4}},{\psi _R} = \psi _{1,4}^{'}{\rm{ = }}{\psi _{5,8}},\\
  or \;{\psi _{\rm{L}}} = \psi _{1,4}^{'},{\psi _R} = {\psi _{1,4}}{\rm{ = }}{\psi _{5,8}}.
\end{array}
\end{split}
\end{equation}

\textbf{Case (1)}\\

The initial state is also a product state of two entangled QD pairs
\begin{equation}
\label{equ12}
\begin{split}
%\begin{array}{l}
\phi
&= {\psi _{1,4}}\otimes {\psi _{5,8}} \\
&= \frac{1}{2}[{\left| {ge} \right\rangle _{1,8}}{\left| {eg} \right\rangle _{4,5}} - i{\left| {gg} \right\rangle _{1,8}}{\left| {ee} \right\rangle _{4,5}}\\
&\quad- i{\left| {ee} \right\rangle _{1,8}}{\left| {gg} \right\rangle _{4,5}} - {\left| {eg} \right\rangle _{1,8}}{\left| {ge} \right\rangle _{5,4}}].
%\end{array}
\end{split}
\end{equation}
By switching on the coupling between the cavity and the QDs 4 and 5, the system evolves after a time $t$ into
\begin{equation}
\label{equ13}
\begin{split}
\phi ^{'}
 &= \frac{1}{2}\{ {\left| {ge} \right\rangle _{1,8}}{e^{ - i\lambda t}}[\cos (\lambda t){\left| {eg} \right\rangle _{4,5}} - i\sin (\lambda t){\left| {ge} \right\rangle _{4,5}}]\\
 &\quad- i{e^{ - it2\lambda }}{\left| {gg} \right\rangle _{1,8}}{\left| {ee} \right\rangle _{4,5}}- i{\left| {ee} \right\rangle _{1,8}}{\left| {gg} \right\rangle _{4,5}} \\
 &\quad- {\left| {eg} \right\rangle _{1,8}}{e^{ - i\lambda t}}[\cos (\lambda t){\left| {ge} \right\rangle _{4,5}} - i\sin (\lambda t){\left| {eg} \right\rangle _{4,5}}]\}.
\end{split}
\end{equation}
Choosing $\lambda t = \pi /4$, we find
\begin{equation}
\label{equ14}
\begin{split}
\phi ^{'}
&= \frac{{{e^{ - \frac{\pi }{4}i}}}}{{2\sqrt 2 }}\left( {{{\left| {ge} \right\rangle }_{1,8}} + i{{\left| {eg} \right\rangle }_{1,8}}} \right){\left| {eg} \right\rangle _{4,5}} \\
&\quad- ({\left| {eg} \right\rangle _{1,8}} + i{\left| {ge} \right\rangle _{1,8}}){\left| {ge} \right\rangle _{4,5}}\\
&\quad- i{e^{ - \frac{\pi }{4}i}}{\left| {gg} \right\rangle _{1,8}}{\left| {ee} \right\rangle _{4,5}} \\
&\quad- \sqrt 2 i{e^{\frac{\pi }{4}i}}{\left| {ee} \right\rangle _{1,8}}{\left| {gg} \right\rangle _{4,5}}.
\end{split}
\end{equation}
Similarly, by detecting the states of QDs 4 and 5, we throw away the results if they are ${\left| {gg} \right\rangle _{4,5}}$ or ${\left| {ee} \right\rangle _{4,5}}$. Otherwise, the final state collapses into the entangled states between QDs 1 and 8, namely
\begin{equation}
\label{equ15}
\begin{split}
\left| \psi  \right\rangle _{1,8 }= \frac{1}{{\sqrt 2 }}({{\left| {ge} \right\rangle }_{1,8}} + i{{\left| {eg} \right\rangle }_{1,8}} ),\\
\left| \psi  \right\rangle _{1,8}^{'} = \frac{1}{{\sqrt 2 }}({\left| {eg} \right\rangle _{1,8}} + i{\left| {ge} \right\rangle _{1,8}}).
\end{split}
\end{equation}
Obviously, the entanglement distance is doubled again.

\textbf{Case (2)}\\

In this case, the initial state is
\begin{eqnarray}
\label{equ16}
         \phi  &=& {\psi' _{1,4}}\otimes{\psi _{5,8}} \nonumber\\
               &=& \frac{1}{2}\big[{{\left| {eg} \right\rangle }_{1,8}}{{\left| {ge} \right\rangle }_{4,5}}- i{{\left| {ee} \right\rangle }_{1,8}}{{\left| {gg} \right\rangle }_{4,5}}\nonumber\\
               &-& i{{\left| {gg} \right\rangle }_{1,8}}{{\left| {ee} \right\rangle }_{4,5}}-{{\left| {ge} \right\rangle }_{1,8}}{{\left| {eg} \right\rangle }_{4,5}}\big].
\end{eqnarray}
By switching on the coupling between the cavity and the QDs 4 and 5, the system evolves after a time $t$ into
\begin{equation}
\label{equ17}
\begin{split}
\phi'
&= \frac{1}{2}\{ {\left| {eg} \right\rangle _{1,8}}{e^{ - i\lambda t}}\left[ {\cos (\lambda t){{\left| {ge} \right\rangle }_{4,5}} - i\sin (\lambda t){{\left| {eg} \right\rangle }_{4,5}}} \right]\\
&\quad- i{\left| {ee} \right\rangle _{1,8}}{\left| {gg} \right\rangle _{4,5}}
 - i{e^{ - i2\lambda t}}{\left| {gg} \right\rangle _{1,8}}{\left| {ee} \right\rangle _{4,5}}\\
&\quad- {\left| {ge} \right\rangle _{1,8}}{e^{ - i\lambda t}}\left[ {\cos (\lambda t){{\left| {eg} \right\rangle }_{4,5}} - i\sin (\lambda t){{\left| {ge} \right\rangle }_{4,5}}} \right]\}.
\end{split}
\end{equation}
Choosing $\lambda t = \pi /4$, we obtain
\begin{equation}
\label{equ18}
\begin{split}
\phi'
&= \frac{{{e^{ - i\frac{\pi }{4}}}}}{{2\sqrt 2 }}\{ ({\left| {eg} \right\rangle _{1,8}} + i{\left| {ge} \right\rangle _{1,8}}){\left| {ge} \right\rangle _{4,5}} - ({\left| {ge} \right\rangle _{1,8}}\\
&\quad+ i{\left| {eg} \right\rangle _{1,8}}){\left| {eg} \right\rangle _{4,5}}- \sqrt 2 i{e^{i\frac{\pi }{4}}}{\left| {ee} \right\rangle _{1,8}}{\left| {gg} \right\rangle _{4,5}}\\
&\quad- i{e^{ - i\frac{\pi }{4}}}\sqrt 2 {\left| {gg} \right\rangle _{1,8}}{\left| {ee} \right\rangle _{4,5}}\}.
\end{split}
\end{equation}
By detecting the states of QDs 4 and 5, we throw away the results if they are ${\left| {gg} \right\rangle _{4,5}}$ or ${\left| {ee} \right\rangle _{4,5}}$. Otherwise, the final state collapses into the entangled states between QDs 1 and 8, namely
\begin{equation}
\begin{split}
\label{equ19}
\begin{array}{l}
\psi _{1,8}^{'} = \frac{1}{{\sqrt 2 }}({\left| {eg} \right\rangle _{1,8}} + i{\left| {ge} \right\rangle _{1,8}}),\\
{\psi _{1,8}} = \frac{1}{{\sqrt 2 }}({\left| {ge} \right\rangle _{1,8}} + i{\left| {eg} \right\rangle _{1,8}}).
\end{array}
\end{split}
\end{equation}

\textbf{Case (3)}

In this case, the initial state is
\begin{eqnarray}
\label{equ20}
     \phi &=&{\psi _{1,4}}\otimes{\psi' _{5,8}}\nonumber\\
      &=&\frac{1}{2}\big[{{\left| {gg} \right\rangle }_{1,8}}{{\left| {ee} \right\rangle }_{4,5}} - i{{\left| {ge} \right\rangle }_{1,8}}{{\left| {eg} \right\rangle }_{4,5}}\nonumber\\
      &-& i{{\left| {eg} \right\rangle }_{1,8}}{{\left| {ge} \right\rangle }_{4,5}} - {{\left| {ee} \right\rangle }_{1,8}}{{\left| {gg} \right\rangle }_{4,5}} \big].
\end{eqnarray}
By switching on the coupling between the cavity and the QDs 4 and 5, the system evolves after a time $t$ into
\begin{equation}
\label{equ21}
\begin{split}
\phi'
&=\frac{1}{2}[{e^{ - i2\lambda t}}{\left| {gg} \right\rangle _{1,8}}{\left| {ee} \right\rangle _{4,5}} \\
&\quad- i{\left| {ge} \right\rangle _{1,8}}{e^{ - \lambda t}}\left[ {\cos (\lambda t){{\left| {eg} \right\rangle }_{4,5}}-i\sin (\lambda t){{\left| {ge} \right\rangle }_{4,5}}} \right]\\
&\quad- i{\left| {eg} \right\rangle _{1,8}}{e^{ - \lambda t}}\left[ {\cos (\lambda t){{\left| {ge} \right\rangle }_{4,5}} - i\sin (\lambda t){{\left| {eg} \right\rangle }_{4,5}}}\right]\\
&\quad- {\left| {ee} \right\rangle _{1,8}}{\left| {gg} \right\rangle _{4,5}}].
\end{split}
\end{equation}
Choosing $\lambda t = \pi/4$, we obtain
\begin{equation}
\label{equ22}
\begin{split}
\phi'
 &= \frac{{{e^{ - i\frac{\pi }{4}}}}}{{2\sqrt 2 }}[\sqrt 2 {e^{ - i\frac{\pi }{4}}}{\left| {gg} \right\rangle _{1,8}}{\left| {ee} \right\rangle _{4,5}} - \sqrt 2 {e^{\frac{\pi }{4}}}{\left| {ee} \right\rangle _{1,8}}{\left| {gg} \right\rangle _{4,5}}\\
 &\quad- ({\left| {eg} \right\rangle _{1,8}} + i{\left| {ge} \right\rangle _{1,8}}){\left| {eg} \right\rangle _{4,5}}\\
 &\quad- ({\left| {ge} \right\rangle _{1,8}} + i{\left| {eg} \right\rangle _{1,8}}){\left| {ge} \right\rangle _{4,5}}].
\end{split}
\end{equation}
By detecting the states of QDs 4 and 5, we throw away the results if they are ${\left| {gg} \right\rangle _{4,5}}$ or ${\left| {ee} \right\rangle _{4,5}}$. Otherwise, the final state collapses into the entangled states between QDs 1 and 8, namely
\begin{equation}
\label{equ23}
\begin{split}
\begin{array}{l}
{\psi _{1,8}} = \frac{1}{{\sqrt 2 }}({\left| {eg} \right\rangle _{1,8}} + i{\left| {ge} \right\rangle _{1,8}}),\\
\psi _{1,8}^{'} = \frac{1}{{\sqrt 2 }}({\left| {ge} \right\rangle _{1,8}} + i{\left| {eg} \right\rangle _{1,8}}).
\end{array}
\end{split}
\end{equation}

By comparing the results of 4 QDs and 8 QDs, we find obviously they are exactly the same, indicating the periodicity of the quantum repeater. It is straightforward to extend the quantum entanglement for 16 QDs with a product state of a pair of 8 entangled QDs~(Fig.\ref{fig:figure3}(c)). Similarly, there are also three cases in the 16 QDs' system
\begin{equation}
\label{equ24}
\begin{split}
\begin{array}{l}
(1).\; {\psi _{\rm{L}}} = {\psi _R} = {\psi _{1,8}},\\
(2).\; {\psi _{\rm{L}}} = {\psi _R} = \psi _{1,8}^{'},\\
(3).\; {\psi _{\rm{L}}} = {\psi _{1,8}},\; {\psi _R} = \psi _{1,8}^{'};\; {\rm or}\; {\psi _{\rm{L}}} = \psi _{1,8}^{'},\; {\psi _R} = {\psi _{1,8}}.
\end{array}
\end{split}
\end{equation}
By repeating this process, it is easy to extend the entanglement distance to 32 QDs and longer.

\begin{figure}[ht]
\centering
\includegraphics[angle=-90,scale=0.30]{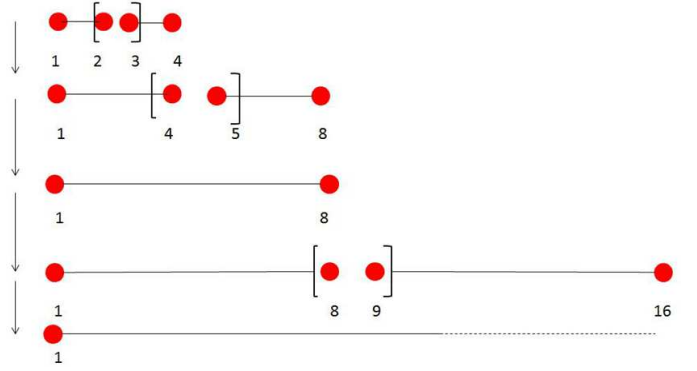}
\caption{A complete process to establish entanglement between QD 1 and QD 16 (and more distant QDs) through coupling two central QDs within cavities.}
\label{fig:figure4}
\end{figure}

Our scheme of a quantum repeater differs from the traditional DLCZ scheme in 3 aspects (see Fig.~\ref{fig:figure4} as a summary). (i) Our system is based on solid state QDs which are different from the coupled atoms and photons system. (ii) The start unit of our scheme is 4 entangled QDs which are prepared from 2 pairs of QDs initially in singlet states. This is different from the atoms and photons system. (iii) The measurements we need in our scheme are local, instead of the Bell measurement required in the DLCZ schemes. These local measurements are easier to implement in experiments than the Bell measurements. Although our scheme of the quantum repeater is different from the traditional one, the success probability of our scheme is the same as the DLCZ scheme, which is higher than other schemes based on solid state systems.

%%%%%%%%%%%%%%%%%%%%%%%%%%%%%%%%%%%%%%%%%%%%%%%%%%%%%%%%%%%%%%%%%%%%%%%%%%%%%%%%%%%%%%%%%%%%%%%%%%%%%%%%
\section{\label{sec:level3}Conclusion}

In conclusion, we propose a quantum repeater scheme based on the solid state QDs. The starting unit is 4 entangled QDs with an experimentally prepared two pairs of DQDs in their singlet states. Our scheme reaches the same success probability as the famous DLCZ scheme but without the requirement of the Bell measurements. Such an advantage is more favored for experimentalists due to the simplicity of the scheme. Quantum repeater is a fundamental block in quantum communication, quantum computing, and quantum teleportation~\cite{Bouwmeester1997Experimental}. For a practical quantum repeater, other factors, such as the decoherence, the environmental effect, the measurement efficiency, need to be included. The performance of the quantum repeater scheme in a real situation and new methods to improve the success probability of entanglement swapping using entanglement purification~\cite{Pan2000Entanglement,PhysRevLett.76.722,PhysRevA.57.R4075} and noise-suppression also require many further explorations.

%%%%%%%%%%%%%%%%%%%%%%%%%%%%%%%%%%%%%%%%%%%%%%%%%%%%%%%%%%%%%%%%%%%%%%%%%%%%%%%%%%%%%%%%%%%%%%%%%%%%%%%%
\section*{ACKNOWLEDGMENT}

X.F.Yi thanks Wenxian Zhang,Feng Mang,Yong Zhang and Zhang-qi Yin for valuable discussions. This work was supported by the National Natural Science Foundation of China under Grant No. 11574239 and Open Research Fund Program of the State Key Laboratory of Low Dimensional Quantum Physics under Grant No. KF201614.

\bibliography{reference}
\end{document}